# CONTINUOUS SIGN LANGUAGE RECOGNITION FROM WEARABLE IMUs USING DEEP CAPSULE NETWORKS AND GAME THEORY

Karush Suri, Rinki Gupta

ABSTRACT

Sign Language is used by the deaf community all over world. Internationally, various research groups are working towards the development of an electronic sign language translator to enhance the accessibility of a signer. By employing intelligent models and wearable devices such as inertial measurement units (IMUs), continuous signs leading to the formation of a complete sentence can be recognized effectively. The work presented here proposes a novel one-dimensional deep capsule network (CapsNet) architecture for continuous Indian Sign Language recognition by means of signals obtained from a custom designed wearable IMU system. The IMU records tri-axial acceleration and turn rate, and orientation of the sensor is evaluated using a complementary filter. All the signals are used in the proposed deep learning network for learning and recognition of the signed sentences. The performance of the proposed CapsNet architecture is assessed by altering dynamic routing between capsule layers. Performance of the model is compared to that of foundational convolutional neural networks (CNNs) in terms of accuracy, loss, false predictions and learnt activations. The proposed CapsNet yields improved accuracy values of 94% (for 3 routing) and 92.50% (for 5 routings) in comparison to CNNs which yield 87.99%. Improved learning of the architecture is also validated by spatial activations depicting excited units at the predictive layer. For the purpose of evaluating relative performance and competitive nature of models, a novel non-cooperative pick game is constructed. The game presents a pick-and-predict competition between CapsNet and CNN constrained to a single strategy adoption. Both models compete with each other in order to reach their best responses. Higher value of Nash equilibrium for CapsNet as compared to CNN indicates the suitability of the proposed approach.

INTRODUCTION

Human motion analysis has been studied for various applications using different sensing technologies [1-9]. For instance, human gait may be analysed to localize the phases in a gait that may be used for health monitoring or navigation purpose [1,2]. The use of an inertial measurement unit (IMU) is very common in human gait analysis because of its low-cost and wearability. An IMU consists of a tri-axial accelerometer and a tri-axial gyroscope that measure the acceleration and turn rate along the $x$-, $y$- and $z$-axis of the sensor. Hence, an IMU provides the capability to assess the motion of an object in three-dimensional cartesian space. Measurement in three-dimensional coordinates also plays a significant role in accurate recognition of hand gestures. Hand-gesture recognition has been applied for designing human-machine interfaces [3, 4]. Again, IMUs are useful for such system because they may be processed to develop real-time systems that could be used to control a computer interface during a presentation [3] or control a virtual-reality head-mounted display based on real-time tracking of upper limb [4].

Hand motion analysis has also been studied for the development of assistive technology such as sign language recognition. Sign language predominantly involves the use of various hand postures and motions. Consequently, different sensing technologies have been employed in analysis of human motion for recognition of the signed word or sentence. These include depth sensors [5], videos [6] as well as wearable sensors such as surface electromyogram (sEMG) and IMU [7,8]. Signs may be recognized from images or video frames by acquisition of the hand posture at an appropriate angle, followed by image segmentation and prediction [5,6]. However, images may not always be suitable for gesture recognition since their acquisition system may not be wearable and images may be affected by illumination condition, foreground focus and background complexity. Hence, wearable sensors may be used for acquiring data while signing. Surface electromyogram measure the muscle potentials when different muscles are activated to perform a specific sign. Several features have been proposed for sEMG signals that may distinguish one hand activity from another [8]. Although they are non-invasive and provide complementary capability as compared to IMUs for hand gesture recognition [7]; sEMG are easily affected by factors such as sensor placement, motion artefacts and subject variability. IMUs estimate the position and orientation of the hand with minimum invasiveness and have a compact design. In this work, an algorithm is proposed for recognition of sentences signed according to the Indian sign language using data from IMU placed in a forearm armband.

While conversing in sign language, a signer performs gestures in a sequence, which result in the formation of a complete sentence. The signs may be evaluated in continuation by means of position and orientation of the hand based on the signals recorded with IMUs. Then, the signal patterns may be analysed by making use of advanced deep models and by exploiting the hierarchical structure of sign language itself to predict the performed sign and hence, translate the signed gestures in the sentence in any verbal language such as English. Recognition of gestures from the recorded signals has been performed using conventional machine learning as well as deep learning techniques [9-10]. Most commonly found algorithms in literature are Artificial Neural Networks (ANNs), which are multilayer perceptrons based on the concept of stacked Restricted Boltzmann Machines (RBMs). ANNs provide a simple framework for layer-based structured classification by making use of feed-forward action and backpropagation algorithm [11]. Introduction of multiple layers in ANN tends to make the network deeper in order to enhance its learning. This gives rise to Deep Neural Networks (DNNs). However, as a result of multiple stacked layers in the structure, ANNs and DNNs suffer from vanishing gradients during training [12]. A suitable alternative to tackle this issue is provided by Recursive Neural Networks (RNNs). RNNs employ recursive layers which can be used time and again for the purpose of classification and generation as well [13]. These are often used to classify sequential data as a result of their recursive nature. Most of the times RNNs are used as generational models which provides higher training times and complex algorithms in collaboration to augment the training process [14]. This generates a need for models capable of improved learning and accurate predictions. Convolution layers for feature learning address this problem. Successive convolution and pooling of data in subsample space augment feature learning, which results in accurate classification. Convolution Neural Networks (CNNs) make use of these layers in a hierarchical manner [15]. Based on two-dimensional convolution, CNNs are excessively used in image processing applications for digit and object recognition [15, 16]. CNN have also been applied on one-dimensional data and shown to yield good classification accuracy of 97.5% [17]. In classification of objects using images, CNNs may provide misclassification when the orientation of the image is altered or when there is a deviation from the expected spatial structure of pixels [18]. Thus, one must aim for equi-variance by making use of Capsule Networks (CapsNet) as proposed by Hinton et'al, which make use of nested group of neurons to construct a capsule [19]. CapsNet decreases layers and errors thereby increasing the prediction accuracy by making use of higher dimensional spacing in capsules. Also, dynamic routing between capsules can further be used to enhance recognition [20]. So far, CapsNet has been applied only for image-based recognition and its utility for multiple data types and dimensions is not reported in literature to the best of our knowledge.

Advanced models for classification are useful only if they optimize the performance. Performance of multiple state-of-the-art classifiers may be compared in terms of quantitative parameters such as classification accuracies, number of false predictions and learning rate. Alternatively, interactive games may be designed wherein the classifiers compete with each other to present better performance. For instance, General Adversarial Networks (GANs) optimize their choice of strategy in order to yield better generational and classification results [21]. Similar to GANs, multi-layered structures may comprise of different learning games which allow the model to study the data better as a result of competition [22].

In this work, the problem of sign language translation is addressed by making use of a novel CapsNet architecture consisting of one-dimensional convolution and dynamic routing. Signed sentences comprising of gestures recorded using a custom-designed wearable IMU device are classified using the CapsNet architecture consisting of capsules having dimensions higher than input data. The details of the recorded dataset and the proposed CapsNet architecture are explained in Section II. Performance of the architecture is compared to the foundational CNN having similar trainable hyper-parameters. Furthermore, validation of accurate predictions is obtained by Game Theory consisting of construction of a non-cooperative pick game in which both the models compete with each other to offer a correct prediction of the input sample by picking the correct class, as explained in Section III. The single-strategy competition serves as the basis for independent performance and sample-by-sample comparison allowing both the models to reach their best response. The results obtained with actual data are presented in Section IV and Section V contains the concluding remarks.

# SIGNED SENTENCE RECOGNITION USING CAPSNET

## (a) DATA CORPUS

The dataset of IMU signals was constructed by making use of the GY-80 multiboard and Arduino UNO board, which contains the ATmega328P microcontroller. The GY-80 board consists of tri-axial accelerometer ADXL345 and gyroscope L3G400D, which measure acceleration (in $m/s^2$) and turn rate (in deg/s), respectively at 100 Hz. Both accelerometer and gyroscope have 3 degree-of-freedom each, giving a total of 6 channels of data. Fig. 1a depicts the GY-80 board consisting of IMU instruments. The setup is placed on wearable bands including straps. This is done for convenient usage of the apparatus and a compact design which would prevent any hindrances during hand motion. The setup is placed on the frontal side of the forearm, below the elbow as depicted in Fig. 1b. The sensors were calibrated for bias and scale factor. Accelerometer data was corrected by using the 12-parameter estimation method [23]. On the other hand, rotational signals were corrected by subtracting bias values and then scaling the sequence [24].

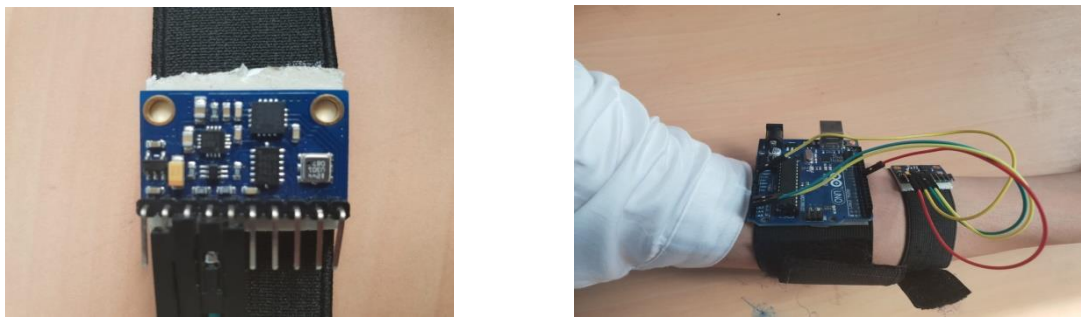

Fig 1. (a) GY-80 sensor board, (b) Placement of experimental apparatus

Signals from IMU sensors were recorded from a total of 10 different subjects out of which 5 are female and 5 are male. All the subjects fall in the age range of 21-49 years and 2 subjects were left-handed which makes the constructed dataset heterogeneous with respect to subject's age, gender and dominant arms. Signs were recorded from the Indian Sign Language (ISL) in a sequence, resulting in the formation of a complete sentence [25]. A total of 20 sentences were gathered with each sentence consisting of 2-4 signs spaced approximately 1s apart. Each subject performed 10 repetitions of a sentence. In ISL, sentences follow a predefined structure. Interrogative words, number quantities and negations are generally signed at the end of the sentence whereas subjects and objects are signed at the beginning. For instance, for the sentence 'I want water', the gesture corresponding to 'I' is performed first followed by gesture corresponding to 'want' followed by gesture corresponding to 'water'. However, in the interrogative sentence 'Where is the Clinic?', gesture corresponding to 'Clinic' is performed first followed by gesture corresponding to 'Where'.

The recorded signals are further processed using a moving median filter to remove any spurious overshoots in the signals introduced because of the recording setup. Euler Angles '$E$' are obtained from accelerometer and gyroscope using (1-2) and (3), respectively, as

$$A_\alpha = atan \frac{A_x}{\sqrt{A_y^2 + A_z^2}}, \quad (1)$$

$$A_\rho = atan \frac{A_y}{\sqrt{A_x^2 + A_z^2}}, \quad (2)$$

and

$$G(\alpha, \rho, \theta) = \int_0^T G(x, y, z) dt. \quad (3)$$

In (1) and (2), the roll ($A_\alpha$) and pitch ($A_\rho$) angles are estimated using accelerations, $A_x$, $A_y$ and $A_z$ measured along $x$-, $y$- and $z$-axes, respectively. In (3), $G(\alpha,\rho,\theta)$ indicates the Euler angles estimated from turn rates $G(x,y,z)$ and '$dt$' indicates the sampling interval and '$T$' denotes the total duration of time. The Euler angles evaluated from accelerometer and gyroscope are combined in a complementary filter to evaluate the final angles as

$$E = \beta * G(\alpha, \rho, \theta) + (1 - \beta) * A(\alpha, \rho, \theta). \quad (4)$$

Here, '$\beta$' indicates the filter coefficient which is selected empirically as 0.85. Euler angles along with the pre-processed accelerometer and gyroscope signals are used in the CapsNet algorithm for classification, as explained below.

### (b) CAPSNET ARCHITECTURE

The use of capsule theory addresses the need for accurate learning algorithms with a fool-proof structure. Two dimensional CaspNet eliminates the problem of rotational confusion present in CNN wherein ambiguous spatial relationships result in a poor performance of the model. In the case of one-dimensional convolution, capsule theory aids in providing a better mapping between the convolved feature set. Nesting of several convolutional layers within a layer gives rise to vectorized structures known as capsules. These capsules act as higher dimensional entities during feature learning by making use of non-linear activations. Unlike the foundational CNNs, excessive convolution takes place in the layers and the need for pooling the outputs into a sub-sample space is prevented. Fig 2 (a) provides the architecture for the one-dimensional CapsNet used in the recognition of IMU signals in contrast to the architecture of CNN which is depicted in Fig 2 (b). In the CapsNet model, convolution of samples at nested primary capsule layer is followed by the Digit capsule (caps) layer. Digit caps layer acts as the higher dimensional feature space for the convolved values to be learnt. The digit caps layer accepts inputs from all the capsules in the previous layer. Non-linear activations at both the primary and digit caps layer are provided by means of the squash function as proposed in the original work on CapsNet [19]. Connections between these two layers are dynamic and are governed by the usage of dynamic routing. Based on the number and intensity of units excited in the next layer, data is routed for learning. This process of routing is governed by logits or simply probabilities denoting the coupling of two successive capsules between two consecutive layers. The coupling coefficients are determined iteratively by making use of softmax, also referred to as 'softmax routing'. These coefficients are refined over iterations in order to provide agreements which are used to link capsules in the previous layer to higher order capsules in the next layer. A conventional network also consists of a mask which is used in the reconstruction of the target vector. However, in the case of gesture recognition sequence generation is not the primary objective and would only add to training time. Thus, the need for a mask is eliminated and a fully connected layer is used for dimensionality reduction and prediction.

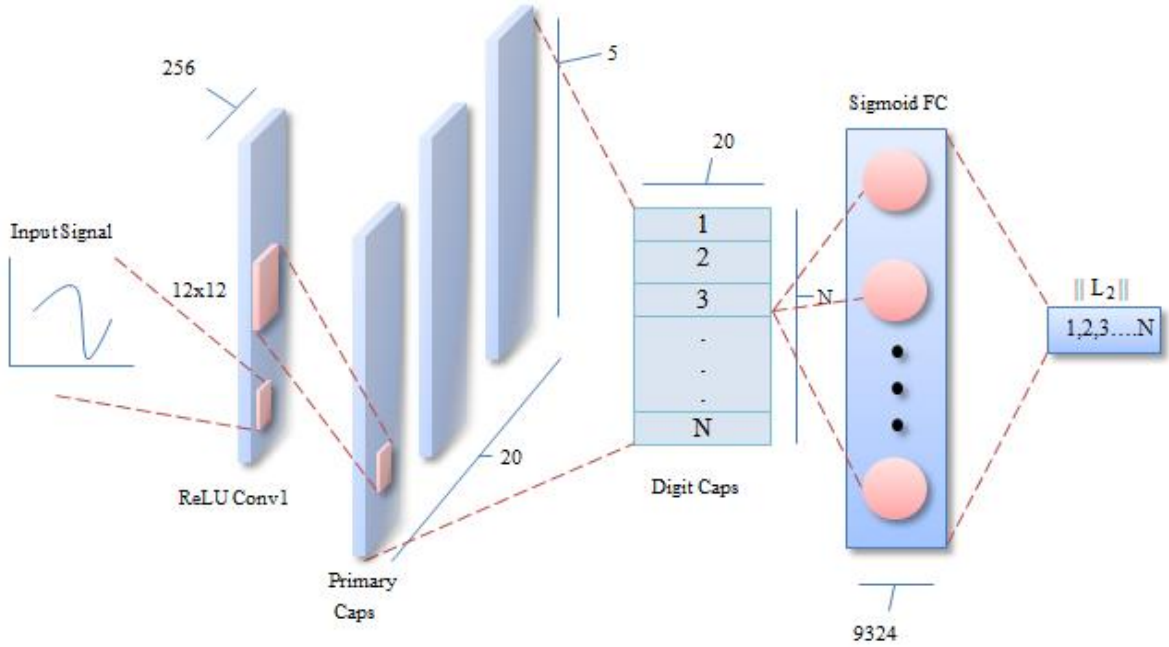

Fig 2 (a) One-Dimensional CapsNet Architecture for Signed Sentence Recognition

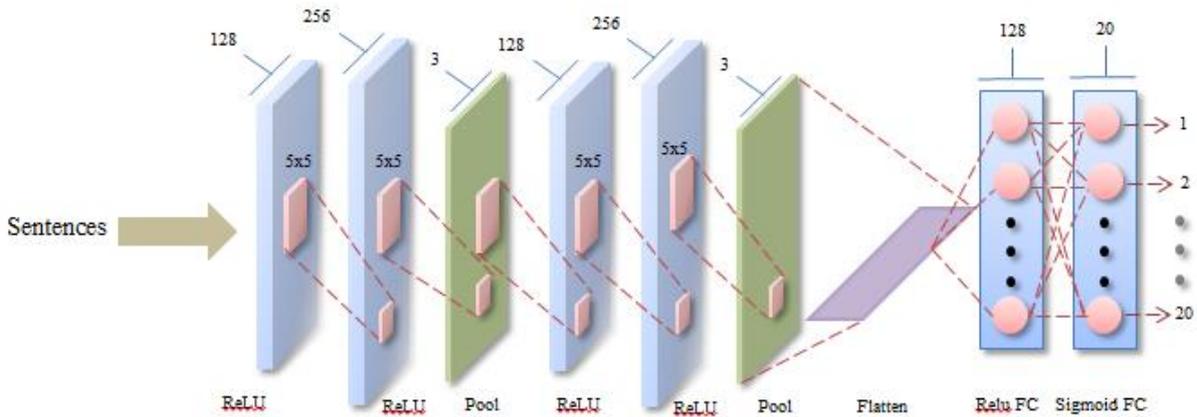

Fig 2 (b) One-Dimensional CNN Architecture

The CapsNet architecture used in this work for the recognition of IMU signals consists of a 12x12 convolutional filter for both the capsule layers with a stride of 1. A total of 256 filters have been used for convolution. The Primary Caps layer has 20 channels with 5 dimensional capsules indicating that each capsule consists of 5 convolutional filters of size 12x12 with a stride of 2. Here, the stride value is increased to 2 keeping in mind the increasing number of convolution coefficients being delivered from the previous layer. Each primary capsule receives the input of all 256x144 1st convolutional layer (Conv1) units. Both the layers are activated with rectified linear unit (ReLU) activation. The ReLU activation has been selected empirically. The Digit Caps layer receives 20x1 capsule outputs from the Primary Caps layer as 5 dimensional vectors. Each capsule in the Digit Caps grid shares its weight with other capsules. A total of 10 dimensions are accumulated per class in the Digit Caps layer. In the case of routing, all the logits are initialized to zero, thus indicating equal probability for the capsule output to be routed to the next unit. For prediction purpose a fully connected Sigmoid layer consisting of 9324 units is inserted. One-dimensional scalar outputs from the Digit Caps layer are received by the fully connected units which lead to the prediction of the correct class. Optimization of the model is conducted by making use of the Adam optimizer with Amsgrad gradient optimization [26]. For a fixed learning rate of 0.001,

Adam provides faster and efficient optimization in comparison to Sigmoid Gradient Descent (SGD) and RMSProp optimizers.

The CNN architecture depicted in Fig 2 (b) is used for comparing the performance of the proposed CapsNet architecture. As shown in Fig 2(b), a filter size of 5x5 has been used because it provides the best precision or the considered dataset. Successive convolution layers consisting of 128 and 256 filters have been used which extract 2000x128 and 128x256 feature values respectively. These values are then accommodated in a subsample pool with a pool size of 3. Reducing the pool size helps the model preserve information in a reduced dimensional space such that it can be readily extracted and delivered to the next layer. Hence, the subsample pool size is selected as 3. Similarly, the 128 and 256 convolutional filters in the convolutional layers are followed by a subsample pool of pool size 3. A flattening layer has been added in order to reduce the dimensionality of the incoming information values and make them compatible with the fully connected perceptrons [17]. A total of 2000 samples of observations are fed to the fully connected layers consisting of 128 and 12 hidden units. The first layer has an excitation of ReLU while the output layer has a sigmoid activation in order to quantize the predictions in accordance with the class values. Optimization and loss parameters of the CNN model are kept same as that of the CapsNet architecture for a uniform basis of comparison.

The performance of the proposed CapsNet algorithm is assessed in terms of quantitative measures such as classification accuracies, variation of loss function and the number of false predictions and the results are compared to the conventional CNN classifier. Moreover, the concept of game theory is utilized to build a non-cooperative pick game to compare the performance of the proposed CapsNet with the conventional CNN classifier, as explained in the following section.

## NON-COOPERATIVE CAPSNET GAMES

Making use of validation techniques has always been good practice for comparing different types of models. Custom methods suitable to the environment and models are preferred in comparison to the conventional methods [27]. Optimized models may be compared using competitive or non-competitive games [28]. Here, a non-cooperative pick game is constructed for comparing and validating the performance of different CapsNet architectures and the conventional CNN. The proposed non-cooperative pick game is designed using the following parameters:

*Players:* CapsNet and CNN architectures are selected as the primary players for the game. However, altering the structural parameters of the CapsNet leads to the introduction of new players in the competition.

*Strategies:* In a standard non-cooperative game, each player can play more than one strategy and optimize its selection of the best strategy for performance. For the considered models, these strategies can be distinguished on the basis of hyper-parameters. However, for the purpose of uniform evaluation and consistent optimization in the given set of parameters, each player is constrained to play only one single pre-defined strategy on the basis of its optimizer.

*Actions:* Actions consist of the set of rules and steps that govern the procedure of the game. Players engaging in the non-cooperative game interact with each other by means of these steps. For the proposed non-cooperative pick game, actions of players are defined on the basis of the following three theorems-

*Theorem-1:* The competing model learns and optimizes upon set $(D_i)^{train}$ by yielding predictions $c_i$ corresponding to each sample in $(D_i)^{train}$ where $c_i \epsilon R$ and $i = \{1,2,....,n\}$.

Here, $(D_i)^{train}$ is the set of all the samples to be used for training of the model (hereby referred to as the training dataset), '$R$' is the set of classes from which model will pick the most appropriate value and '$n$' is the number of samples in the training dataset. Optimization of the model is carried out during the training phase wherein the model picks most appropriate class corresponding to each sample in the training dataset. Successive iterations of this process result in the learnt weights which indicate the peak performance of the model.

*Theorem-2: The competing model predicts upon set $(D_i)^{test}$ by yielding predictions $c_j$ corresponding to each sample in $(D_i)^{test}$ where $c_j \in R$ and $j = \{1,2,....,m\}$.*

Here, $(D_i)^{test}$ is the set of all the samples to be used for testing of the model (hereby referred to as the testing dataset) and '$m$' is the number of samples in the testing dataset. The model makes predictions on the testing samples by picking a class from the set of classes once its performance has been optimized.

*Theorem-3: The model is said to have won the competition if $c_j = c_{true}$ and for its competitor $c_j \neq c_{true}$ where $c_j$, $c_{true} \in R$. In any other case the competition is declared to be a draw.*

The above theorem clearly indicates the rules for deciding the winner of the game. If exactly one of the competing models picks the correct class, i.e.- the selected class is same as the true class then that model is said to have won the competition. In any other cases such as multiple models selecting the correct class or none of the models selecting the correct class the game concludes in a draw between all the models. If there are '$x$' players competing in a game then the condition '$P$' for a player to win the game is mathematically expressed by Eq. 5.

$$P(1): (c_j)_1 = c_{true} ; P(2): (c_j)_2 = c_{true}; \ldots \ldots \ldots P(x): (c_j)_x = c_{true} \quad (5)$$

These can be grouped together in a set denoted by $W$ represented by Eq. 6.

$$W = \{P(1), P(2), \ldots \ldots P(x)\} \quad (6)$$

Now, for a player '$z$' to win the game, as depicted in Eq. 7.

$$\exists! z: W(z) ; \; z \in \{1,2,. \ldots x\} \quad (7)$$

Thus, there is exactly one player '$z$' in the set of possible winning criteria '$W$' which satisfies the condition. However, the conditions for the game to end up as a draw is expressed by Eq. 8, 9 and 10.

$$\nexists z: W(z) ; \; z \in \{1,2,. \ldots x\} \quad (8)$$

$$\vee z: W(z) ; \; z \in \{1,2,. \ldots x\} \quad (9)$$

$$\vee! z: W(z) ; \; z \in \{1,2,. \ldots x\} \quad (10)$$

These are combined together and represented by Eq. 11.

$$\nexists z \;|\; \vee z \;|\; \vee! z : W(z) ; \; z \in \{1,2,. \ldots x\} \quad (11)$$

The game comes out to be a draw if all or none or more than one player in the competition pick the correct class. Thus, winning of the constructed non-cooperative pick game is an exclusive task which depends not only on the behaviour of the model's performance but also on the probability that the competing models fail.

---

Algorithm 1.

for *i*=1:*n*
    $(c_i)_{CNN}$ ← pred($(D_i)^{train}$)
    $(c_i)_{CapsNet}$ ← pred($(D_i)^{train}$)
for *j*=1:*m*
    $(c_j)_{CNN}$ ← pred($(D_i)^{test}$)
    $(c_j)_{CapsNet}$ ← pred($(D_i)^{test}$)
    if $((c_j)_{CapsNet} == (c_j)_{true}$ & $(c_j)_{CNN} \neq (c_j)_{true})$
        Winner ← CapsNet
    else if $((c_j)_{CNN} == (c_j)_{true}$ & $(c_j)_{CapsNet} \neq (c_j)_{true})$
        Winner ← CNN

Algorithm 1 highlights the procedure for the constructed novel non-cooperative pick game. It makes use of the training and validation phases of both the networks which act as players to the constructed game. A total '$n$' number of predictions are made by iterating over all the samples in the training dataset by making use of the predict (pred) procedure. The procedure is a simple function consisting of prediction operations of the networks. Each model selects the most appropriate class as per its strategy on the basis of learnt weights. $(c_i)_{CNN}$ denotes the $i^{th}$ prediction offered by CNN and $(c_i)_{CapsNet}$ denotes the $i^{th}$ prediction offered by CapsNet. During the validation phase, the operation of the game begins by making a total '$m$' predictions on the testing dataset. Over each iteration these predictions are compared with the $(c_j)_{true}$ which indicates the true value of the class corresponding to $i^{th}$ sample. In case the picked class satisfies the winning criteria then the corresponding model is declared as the winner. In any other case the game results in a draw. Since a total of '$m$' iterations lead to '$m$' assessments of the winning criteria, a total '$m$' number of games are utilized to compare the relative performances of CNN and CapsNet with respect to each other.

## RESULTS AND DISCUSSION

### (a) CAPSNET RECOGNITION

The proposed one-dimensional CapsNet model is applied on the IMU signals to perform classification of the signed sentences. The performance of the proposed model is compared to the conventional state-of-the-art CNN having similar hyper-parameters. Both the models make use of the same loss function, categorical cross-entropy. However, final layer activation for CNN is empirically determined to be softmax, since it yields better learn weights over each successive iteration. Also, different number of routings are used in the CapsNet architecture to assess the variation of performance of the classifier when the number of dynamic connections is increased or decreased. Fig 3a and Fig 3b depict the training and validation accuracies of the models over 50 iterations respectively. Learning of CapsNet exceeds that of CNN as a result of faster response in improvement over the learnt weights. As observed in Fig. 3, while the training accuracies for 3 or 5 routings overlap, during validation, the model architecture with 3 routings performs better in comparison to 5 connections and CNN. Peak accuracy values for the CapsNet architecture with 3 routings are observed to be 99.72% during training and 94.00% during validation. Figs. 4a and 4b highlight the same variation by means of optimization loss during training and validation, respectively. CapsNet with 3 routings depicts better performance with approximately 0.01 units of loss during training and validation. Another informative measure to assess the predictive behaviour of these models is the number of false predictions. A low value of optimization loss need not necessarily indicate accurate predictions over all the batches. Since loss is calculated as the average over all the batches, misclassification of values between batches tends to be neglected in this process. Figs. 5a and 5b indicate the number of false predictions over each iteration during training and validation, respectively. Fewer false predictions in the case of 3 routings architecture are observed (8 during training and 12 during validation) as compared to false predictions for 5 connections and CNN.

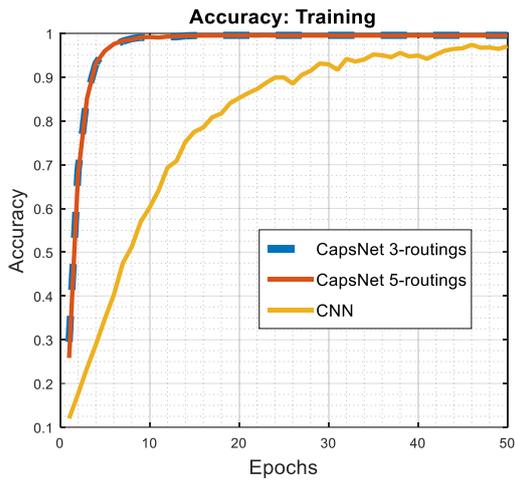
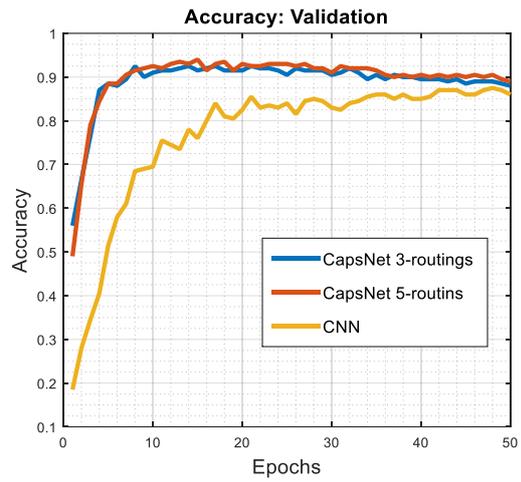

Fig 3. Average Accuracy value variation over 50 iterations for (a) Training and (b) Validation

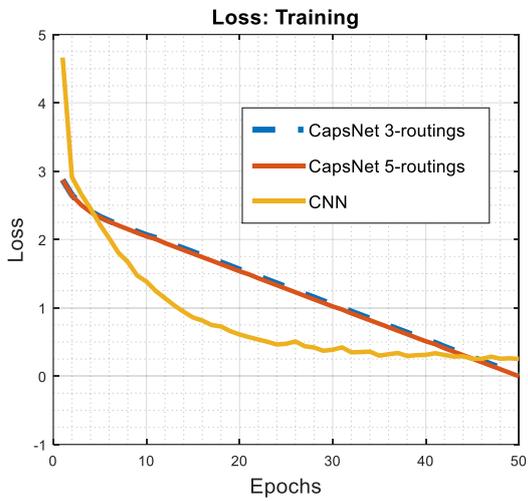
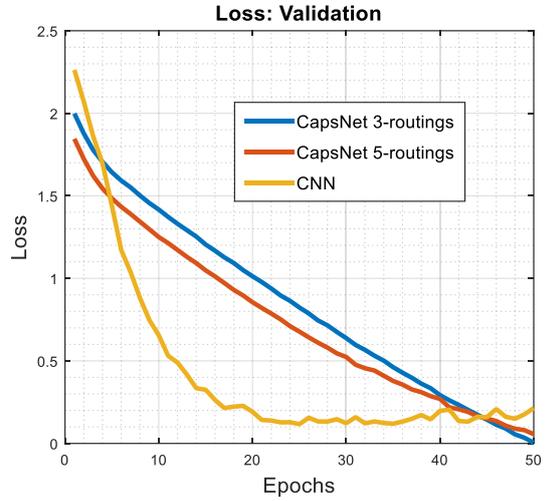

Fig 4. Optimization loss variation (categorical crossentropy) over 50 iterations for (a) Training and (b) Validation

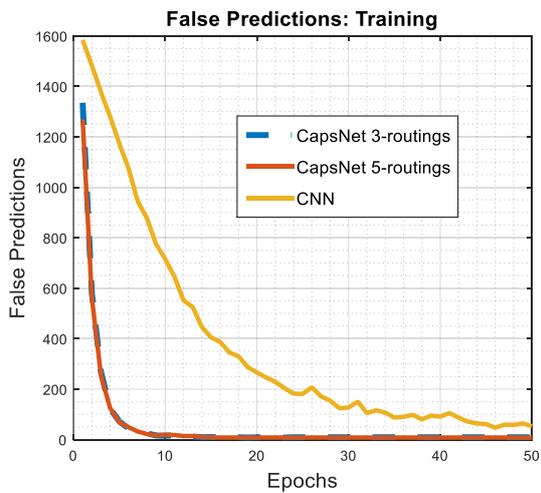
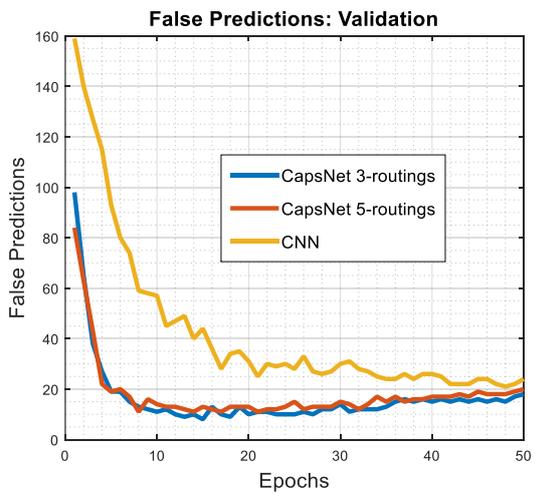

Fig 5. False Predictions variation over 50 iterations for (a) Training and (b) Validation

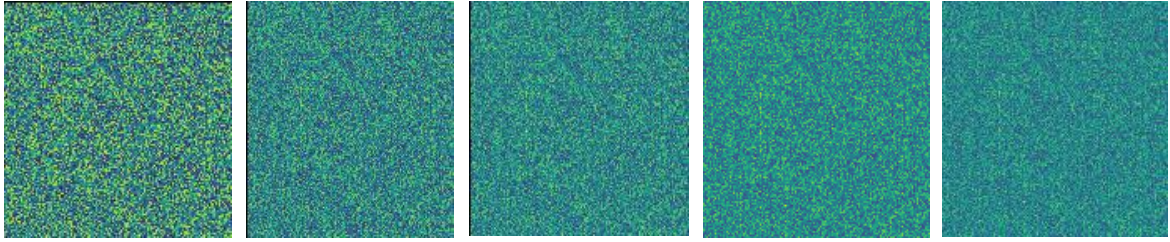

Fig 6. Learnt Activations at the final layer of CNN over (a) 1 iteration, (b) 10 iterations, (c) 20 iterations, (d) 30 iterations and (e) 40 iterations

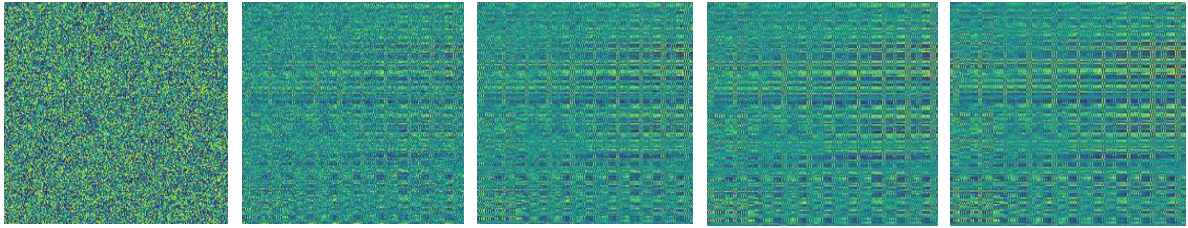

Fig 7. Learnt Activations at the final layer of CapsNet over (a) 1 iteration, (b) 10 iterations, (c) 20 iterations, (d) 30 iterations and (e) 40 iterations

Analysing learnt activations for the model plays a significant role in the evaluation of learning as these weights tend to change over each iteration [29]. Fig. 6 and 7 indicate the improvement in weights after every 10$^{th}$ iteration for CNN and CapsNet architecture. Activation of the last layer tends to improve by virtue of the learnt weights. Higher intensity of the colour gradient depicts highly activated units whereas lower gradients indicate lower values. As seen from Fig. 7a to 7e, the multiplicative values modify over passing iterations producing a regular pattern in Fig. 7e, which depicts the weights in the spatial domain and a definite structure in the background for CapsNet architecture. This definite structure in Fig. 7 (e) indicates the activated units in the layer. However, no such regular pattern is observed to emerge in the case of CNN (Fig. 6a to 6e). Thus, sharp spatial activations with a higher intensity of the colour gradient in Fig. 7 indicate better optimized weights of CapsNet in comparison to CNN. Accurate predictions are a cause of modified activations at the end of the training phase. Comparing weights of CapsNet to CNN, the CapsNet architecture is expected to yield accurate predictions and enhanced learning as a result of the highly activated units arising from dynamic routing between nested units. This is not observed in a conventional CNN due to the absence of nested units and a routing technique.

## (b) NON-COOPERATIVE GAMES

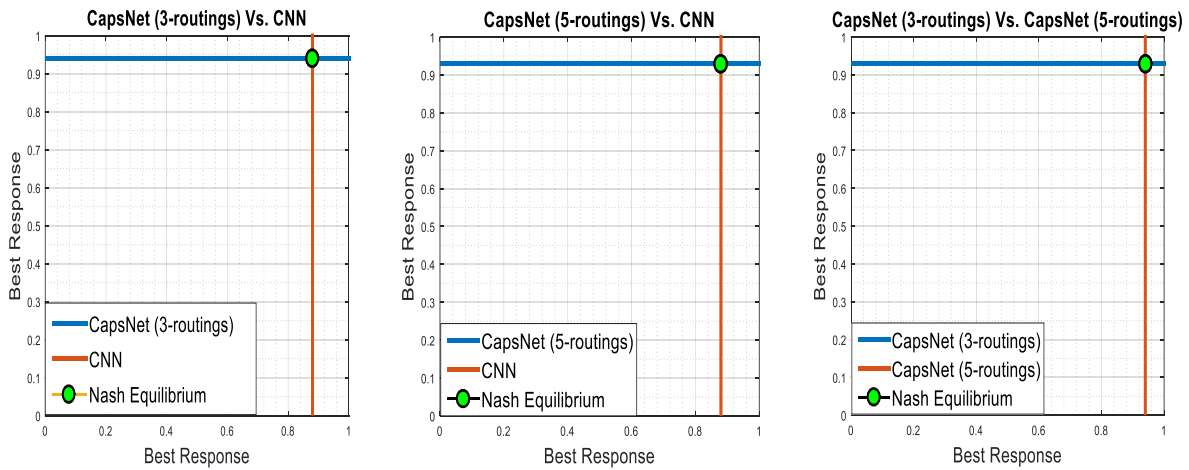

Fig 8. Nash Equilibrium obtained between best responses for (a) CNN Vs. CapsNet (3 routings), (b) CNN Vs. CapsNet (5 routings) and (c) CapsNet (5 routings) Vs. CapsNet (3 routings).

The proposed two-player game designed as explained in Section III, allows the participating models to choose the most appropriate class from the set of all classes based on the information learnt during training. Both the models optimize their performance by learning the values of the weights in order to perform correct predictions. Interactions between players are limited to the extent that they compete with each other during each validation phase. A total of 3 games are played, each between two sets of players at a given time. A simple three-player game would also be equivalent to these games. However, such a scenario would not provide details on the relative performance and would require the use of three-dimensional space for visualization. Each game of the three games consist of the two players. These games based on their players may be mathematically summarised as,

$$\text{Game-1- } P(CNN): (c_j)_{CNN} = c_{true} ; P(CapsNet\text{-}3): (c_j)_{CapsNet\text{-}3} = c_{true} \qquad (12)$$

$$\text{Game-2- } P(CNN): (c_j)_{CNN} = c_{true} ; P(CapsNet\text{-}5): (c_j)_{CapsNet\text{-}5} = c_{true} \qquad (13)$$

$$\text{Game-3- } P(CapsNet\text{-}3): (c_j)_{CapsNet\text{-}3} = c_{true} ; P(CapsNet\text{-}5): (c_j)_{CapsNet\text{-}5} = c_{true} \qquad (14)$$

where $(c_j)_{CNN}$ denotes the class picked by CNN corresponding to the $j^{th}$ sample in the validation set, $(c_j)_{CapsNet\text{-}3}$ denotes the class picked by CapsNet with 3-routings corresponding to the $j^{th}$ sample in the validation set and $(c_j)_{CapsNet\text{-}5}$ denotes the class picked by CapsNet with 5-routings corresponding to the $j^{th}$ sample in the validation set. Collective representation of these games in set notation is expressed by Eq. 15.

$$W_1 = \{P(CNN), P(CapsNet\text{-}3)\}; W_2 = \{P(CNN), P(CapsNet\text{-}5)\}; W_3 = \{P(CapsNet\text{-}3), P(CapsNet\text{-}5)\} \qquad (15)$$

where $W_1$, $W_2$ and $W_3$ indicate games 1, 2 and 3, respectively. Once both the models in a game have reached their best responses, their relative performances may be evaluated by means of Nash Equilibrium which is defined as followed [30],

*Definition-1 (Nash Equilibrium)- Nash Equilibrium is defined as a stable state involving the interaction of different players, in which no player can gain a unilateral change of strategy if the strategies of other players remain unchanged.*

Definition-1 indicates that once the strategy of a player is fixed, it cannot be altered with respect to the strategy of the other player. The idea of Nash Equilibrium corresponds to the point where both the players in the constructed games have adopted and optimized the single selected strategy. It denotes that point in the game where both the players are at their best with respect to each other's performance. Nash Equilibrium is an essential indicator in determining the extent of optimization between the strategies. Once the strategies have been optimized the models are said to have their best responses during the interaction. In this work, the players are CNN and CapsNet with 3- or 5-routings. The strategy of the players is the optimization of the weights that happens during the training phase. The pick games are played during the testing phase. Relative performance between the two competing models in games is thus assessed by visualizing their best responses. Fig 8 indicates the best responses for both the models in games with respect to each other. Best response here refers to the probability that the class selected by the model is same as the true class. The point of intersection on the best response plot denotes the Nash Equilibrium for the corresponding game. Both the players have adopted and optimized their single selected strategy in order to reach the point of Nash Equilibrium. Of the two models CNN and CapsNet, the CapsNet architecture depicts a higher probability of winning in comparison to CNN. CapsNet with 3-routings achieves the best performance in all of its games with a response of 0.94 when compared to CapsNet with 5-routings and CNN that achieve their best performance with responses of 0.93 and 0.88, respectively.

Now, the computational complexity and time required by the considered models is compared. Each iteration of the training phase takes 180 s for the conventional CNN, 480s ± 1μs for CapsNet with 3-routings and 520s for CapsNet with 5-routings. Although, the time required by CapsNet for optimization is higher than CNN, however during validation, CapsNet predicts values faster and more accurately than CNN as a result of increased weight optimization. Since optimization is absent in the validation phase, it only takes 9ms (45μs per sample) and 10ms (50μs per sample) for CapsNet with 3- and 5- routings to make the predictions, respectively. Thus, each game, in the case of 3-routings, takes 45μs to complete as result of the models being pre-trained.

Table 1. Performance Comparison of CapsNet architecture with CNN

| Architecture | Optimization Loss | | Misclassified Outputs | | Classification Accuracy | | Nash Equilibrium | |
| --- | --- | --- | --- | --- | --- | --- | --- | --- |
| | Training | Validation | Training | Validation | Training | Validation | Training | Validation |
| CNN | 0.07 | 0.11 | 139 | 24 | 93.00% | 87.99% | 0.93 | 0.88 |
| CapsNet 3 routings | **0.01** | **0.01** | **8** | **12** | **99.72%** | **94.00%** | **0.99** | **0.94** |
| CapsNet 5 routings | 0.01 | 0.02 | 8 | 15 | 99.56% | 92.50% | 0.99 | 0.93 |

Table 1 summarizes the performance of CapsNet models in comparison to CNN. CapsNet with 3-routings achieves the highest performance for recognition of the signed sentences from the IMU signals with an optimized loss of 0.01, false predictions as 12, accuracy of 94% and Nash Equilibrium for non-cooperative games at 0.94. CapsNet with 5-routings achieves a 4.50% improvement in recognition and non-cooperative games when compared to CNN. Improved results of the CapsNet architecture in comparison to the conventional CNN highlight the appropriateness of dynamic routing with nested layers in the capsule theory. In the future, these capsules can be further modified as stages in a hierarchical network such as master-slave models and parallel architectures. This would help solve problems related to under-fitting.

CONCLUSION

Sign language is the primary language used by the deaf community. In signing, the signer mainly uses different hand postures and motion to convey his message. Development of an electronic sign language translator to translate signing into verbal languages has gained significant importance in the research community over the past decades. Hand motion analysis using wearable systems for measuring muscle intensity, hand orientation, motion and position is only one part of the requirement of a sign language recognition system. The system also requires proportionally efficient and accurate recognition algorithms capable of recognizing signs continuously with minimum delay.

In this work, recognition of sentences signed according to the Indian sign language is performed using signals recorded from a custom-designed wearable IMU device. The wearable IMU device consists of a tri-axial accelerometer and a tri-axial gyroscope. Additionally, the orientation of the hand is also estimated. The database of IMU signals is recorded for with 10 subjects who performed 10 repetitions of each of the 20 sentences, 12 of which are assertive statements and 8 are interrogative questions. The sentence recognition is carried out using a novel one-dimensional CapsNet architecture. The proposed model makes use of Capsule theory consisting of nested convolutional layers linked to each other by dynamic routing. Information is passed by iteratively selecting the most activated units in the following layers. Performance of the model is assessed when 3 or 5 dynamic routings are used in the CapsNet architecture. The performance of the proposed CapsNet architecture is compared with its predecessor model, a conventional CNN. The models have approximately similar number of convolutional levels and their hyper-parameter values have been kept same in order to yield unbiased results. Comparisons have been drawn for both the networks in terms of quantitative measures such as classification accuracy, evolution of loss function and number of false predictions. All of the measures have been assessed and compared for 50 epochs in the training and validation phases.

Improved classification accuracy of 94% is observed for CapsNet with 3-routings in comparison to CapsNet with 5-routings and CNN, which yield 92.5% and 87.99% classification accuracy, respectively. With a higher recognition rate, CapsNet with 3 iterative routings is declared to be more accurate in comparison to CapsNet with 5 routings and the CNN model. In terms of recognition loss, the 3-routing model of CapsNet has a minimized loss of 0.01 in comparison to 0.02 and 0.11 for the 5 routing CapsNet model and CNN respectively. Thus, CapsNet with 3-routings is better optimized in contrast to the other two models. Learning of the architecture is validated by observing spatial activations depicting excited units at the final layer. The CapsNet model presents a uniform pattern of activated convolutional perceptrons in its spatial weight activations. This is found to be absent in the case of CNN, thus indicating less excitation. Furthermore, a non-cooperative pick game is constructed for assessing the relative performance of the models. The game is constrained to single strategy adoption which is optimized by means of optimization function of the model. Each model can adopt only one optimization function, which is its strategy in the game. Both the models optimize their performances with respect to each other in order to attain Nash Equilibrium. The model with higher recognition accuracy at its Nash Equilibrium is declared to be more accurate in the competition. CapsNet architecture presents a better value of the best response at Nash Equilibrium asserting the suitability of the proposed approach.

## Acknowledgment

The author would like to thank the volunteers who helped in recording the data. The author also recognizes the funding support provided by the Science & Engineering Research Board, a statutory body of the Department of Science & Technology (DST), Government of India (ECR/2016/000637).

VITAE

Karush Suri

Mr. Karush Suri is pursuing his B.Tech in Electronics and Communication Engineering from Amity University, Uttar Pradesh, India. His research is concerned with the application of deep learning and machine intelligence for task specific learning such as hand motion analysis and gesture recognition. Karush has participated in various international conferences and presented his work on Master-Slave Neural Networks and Convolutional Array for sign language recognition. In his final year project thesis titled "Application of Deep Learning and Game Theory for Sign Language Recognition using Wearable Sensors", Karush proposed a novel one-dimensional Capsule Network for recognition of continuously signed sentences in the Indian Sign Language.

Rinki Gupta

Dr. Rinki Gupta received her Ph.D. in Signal Processing from the Centre for Applied Research in Electronics (CARE), Indian Institute of Technology (IIT), Delhi, India in 2014. Thereafter, she was working on a project funded by Ministry of Defence at IIT Delhi. She joined Amity School of Engineering and Technology, Amity University, Uttar Pradesh, India in Nov. 2015. She is also the principal investigator in a project sanctioned by SERB, DST and co-investigator in another project funded by DST. Her research interests include human motion analysis, multi-sensor data fusion and speech and audio signal processing.